\documentclass[a4paper,fleqn,usenatbib]{mnras}

\usepackage[T1]{fontenc}
\usepackage{ae,aecompl}

\usepackage{graphicx}	
\usepackage{amsmath}	
\usepackage{amssymb}	

\title[Solar g$^{(2)}(\tau)$ through Atmospheric Turbulence]{Optical Intensity Interferometry through Atmospheric Turbulence}

\author[P. K. Tan, A. H. Chan and C. Kurtsiefer]{
P. K. Tan$^{1,2}$\thanks{pengkian@physics.org},
A. H. Chan$^{2}$,
and C. Kurtsiefer$^{1,2}$
\\
$^{1}$Centre for Quantum Technologies, 3 Science Drive 2, 117543, Singapore\\
$^{2}$Department of Physics, National University of Singapore, 2 Science Drive 3, 117551, Singapore
}

\pubyear{2015}

\begin{document}
\label{firstpage}
\pagerange{\pageref{firstpage}--\pageref{lastpage}}
\maketitle

\begin{abstract}
Conventional ground-based astronomical observations suffer from image
distortion due to atmospheric turbulence. This can be minimized by choosing
suitable geographic locations or adaptive optical techniques, and avoided
altogether by using orbital platforms outside the atmosphere. One of the
promises 
of optical intensity interferometry is its independence from atmospherically
induced phase fluctuations. By performing narrowband spectral filtering on sunlight and conducting temporal intensity interferometry using actively quenched avalanche photon detectors (APDs), the Solar $g^{(2)}(\tau)$ signature was directly measured. We observe an averaged photon bunching signal of $g^{(2)}(\tau) = 1.693 \pm 0.003$ from the Sun, consistently throughout the day despite fluctuating weather conditions, cloud cover and elevation angle. This demonstrates the robustness of the intensity interferometry technique against atmospheric turbulence and opto-mechanical instabilities, and the feasibility to implement measurement schemes with both large baselines and long integration times.
\end{abstract}




\section{Optical Intensity Interferometry}

\citet{hbt:54, hbt:56} demonstrated that at sufficiently short baselines or timescales, both the spatial and temporal correlation measurements of thermal light sources such as the stars should exhibit a photon bunching signal, which was later generalised by \citet{glauber:63} as the second order correlation, g$^{(2)}$, that peaks at twice the value of the statistically random noise floor:
\begin{equation}\label{eq:g2decay}
	g^{(2)}(\tau, b) = 1 + e^{-2\left|\tau\right|/\tau_{c}}\left|\gamma(b,l_c)\right|^2,
\end{equation}
where $\tau$ is the detection time difference, $\tau_{c}$ the coherence time
of a light source with a Lorentzian spectral profile, and $\gamma$ a (complex)
spatial coherence function depending on the spatial separation $b$ of the two
detectors and an effective spatial coherence length $l_c$ of the light
source. This
correlation signature is independent of the optical phase, and forms the basis for intensity interferometry, which has the benefit of being insensitive to first order noise contribution from urban light pollution and atmospheric turbulence \citep{davis:99}.

The spatial correlation $g^{(2)}(\tau=0,b)$ imparts information about the
shape and intensity distribution of the light source, while the temporal
correlation $g^{(2)}(\tau, b=0)$ reveals the characteristic emission mechanism of its source, such as whether its a coherent laser light or an incoherent thermal source \citep{morgan:66, mandel:95, loudon:00, salehteich:07}.

Intensity interferometric measurements revealing information about the
angular size via $l_c$ \citep{hbt:74}, and the emission nature of the light
source via $t_c$ are so in turn potentially affected by decoherence in the
spatial and temporal domains, respectively. It is therefore important to
investigate the influence of atmospheric effects on intensity correlation
measurement of celestial objects.

\subsection{Atmospheric Turbulence}
Atmospheric turbulence introduces scintillation with a characteristic
timescale on the order of microseconds, and a spatial inner scale of
approximately 3\,mm for typical wind speeds of 10\,ms$^{-1}$ \citep{dravins:97}, corresponding to delay time variations on the order of tens of picoseconds due to fluctuations in the atmospheric path difference \citep{dravins:07, cavazzani:12}. This sets the lower bound on effective detector timing resolution to be in the 10\,ps regime before it is constrained by the atmospheric scintillation. 
This timing uncertainty due to the varying refractive index in the atmospheric layers \citep{marini:73} has been observed in laser ranging experiments \citep{kral:04, blazej:08}. The Wiener-Khinchin theorem asserts that the auto-correlation function of a stationary random process is given by the Fourier transform of its power spectrum, which may pick up this timing uncertainty.

The most commonly considered influence of the atmosphere on astronomical
observations is the seeing, which extends the diffraction-limited Airy
patterns of stellar light sources into seeing discs, with a typical diameter
on the order of arcseconds. 
  

Of more concern for observation of temporal correlations are chromatic
dispersion effects which can be suppressed by operating in a narrow optical
bandwidth. The atmosphere also induces phase difference of about 0.5\,radian between optical light waves with a frequency difference of only a few Gigahertz (GHz) \citep{dravins:05}, and a corresponding loss of correlation estimated to be less than 1\,\%. This suggests a practical spectral bandwidth in the range of GHz to suppress this chromatic dispersion.

In this work, we investigate the influence of turbulence in an atmospheric
column varying over the time of the day on the photon bunching signature in an
intensity correlation measurement, and therefore the effective length of the
air column. The optical bandwidth in our experiment is narrow enough to
resolve the temporal correlation $g^{(2)}(\tau)$ to assess the emission nature
of a light source, and eventually detect decoherence effects of non-atmospheric
origin.

\section{Experimental Setup}
Similar to our earlier work \citep{pk:14}, sunlight is collected by an aspheric lens with an effective focal length of
4.51\,mm into a single mode optical fibre (Thorlabs 460HP,  single mode for 450--600\,nm, numerical aperture NA=0.13). The
small diameter aspheric lens is sufficient here because the Sun is an
angularly extended thermal light source. Thus, the number of photons per spatial mode per unit frequency distribution is defined only by the source surface temperature, and independent of the collection aperture size \citep{stokes:94}.

\begin{figure}
\centering
\includegraphics[width=\hsize]{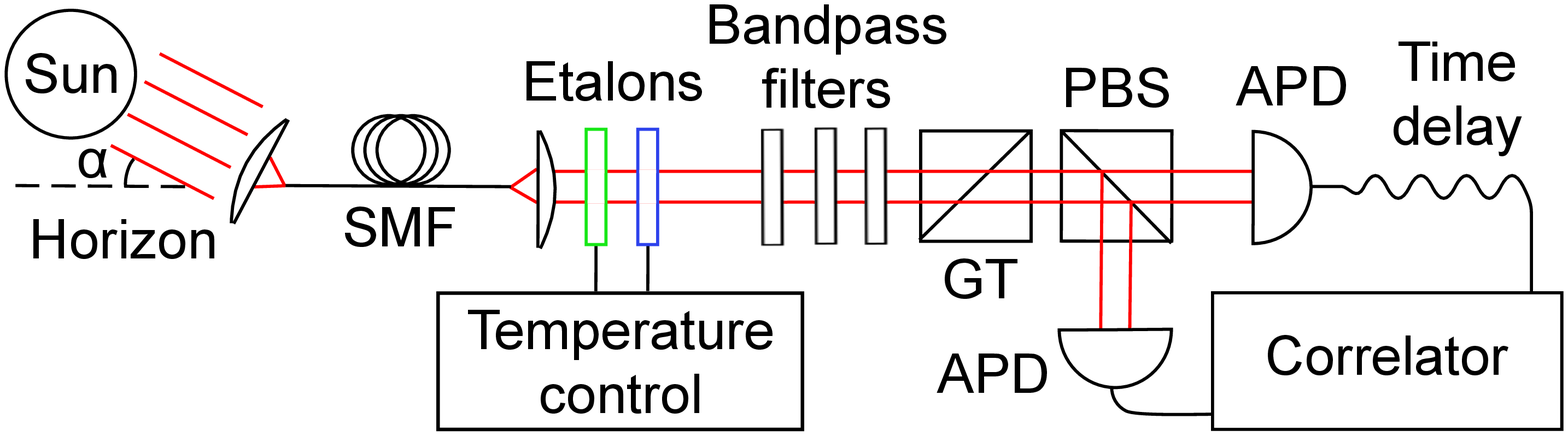}
\caption{Optical setup. Sunlight is coupled into a single mode optical fibre
  for spatial mode filtering, and then exposed to a stack of spectral filters,
  consisting of two temperature-stabilised etalons E1, E2, and a bandpass
  filter stack (BPF) of three interference filters. One polarisation of the
  transmitted narrow spectrum is selected with a Glan-Taylor polariser, and
  distributed with a polarising beam splitter (PBS) onto to avalanche
  photodiodes (APD) for photo-detection time analysis.}
\label{setup}
\end{figure}

The projection into the  fundamental TEM$_{00}$ Gaussian mode of the single mode
fibre leads to a light field with perfect spatial coherence behind the optical
fibre, which is then directed through an arrangement of narrowband spectral
filtering as illustrated in Fig.~\ref{setup}.

\subsection{Filter stack}
In the filtering scheme reported here, Sunlight leaving the optical fibre
is collimated 
by an aspheric lens with an effective focal length of 4.6\,mm. The light is
then sent through two plane-parallel solid etalons made out of fused silica
(Suprasil311) of thickness 0.5\,mm and 0.3\,mm, respectively. With a
refractive index of 
$n=1.460$ around 546--570\,nm, this corresponds to a free spectral range
\begin{equation}
FSR =\frac{c}{2dn}
\end{equation}
of approximately 205\,GHz and 342\,GHz, respectively. The etalons have
reflective coatings of $R$ = 97\,\% at 546.1\,nm on both sides. Neglecting
losses in the coatings, the finesse \citep{fox:06} 
\begin{equation}
{\cal{F_{R}}} = \frac{\pi \sqrt{R}}{1-R} = 103
\end{equation}
of the etalons leads to a transmission bandwidth of $\Delta f_{\rm
  FWHM}=FSR/{\cal{F_{R}}}=3.32$\,GHz for the 0.3\,mm etalon, and $\Delta f_{\rm
  FWHM}=1.99$\,GHz for the 0.5\,mm etalon, respectively.
The spectral transmission profiles of the etalons are tuned via temperature,
with a tuning coefficient of $-4.1$\,\/GHz\,/K. Temperature
tuning was used in lieu of physical rotation of the etalons to avoid
'walk-off' effects \citep{green:80}, and for mechanical stability of the setup.

\begin{figure}
\centering
\includegraphics[width=\hsize]{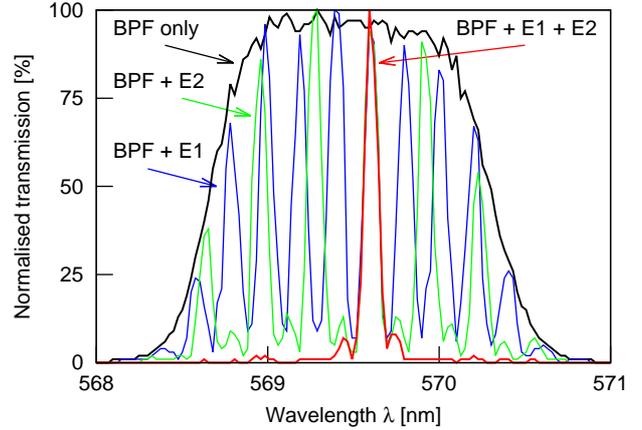}
\caption{Transmission profile of the filter stack. The black trace shows the
  bandpass filter stack (BPF) only, with a width of 2\,nm (FWHM). When adding
  etalon E1 (thickness 0.5\,mm), several transmission peaks fall into the window
  selected by the BPF. Similarly, wider spaced transmission peaks are visible
  with etalon E2 (thickness 0.3\,mm). When combining both etalons and the BPF,
  only one transmission peak is left, with small residual contributions about
  0.6\,nm due to near-overlaps. The line width of the transmission
  peaks  with the etalons is dominated by the spectrometer response of
  0.12\,nm, the actual line width should be around 0.002\,nm. }
\label{stack}
\end{figure}

A series of three interference bandpass filters are used to
reject all other wavelengths: two narrowband filters of central wavelength
(CWL) = 569.6\,nm with Full Width at Half Maximum (FWHM) of 2.2\,nm and a
broadband filter with CWL = 540\,nm and FWHM = 80\,nm. The transmission
profile of the bandpass filters (BPF), and the transmission profiles of a
combination of the BPF with etalons is shown in Fig.~\ref{stack}. For
a single etalon, several transmission peaks corresponding to its
free spectral range fall into the transmission window of the
BPF. By combining both etalons with their different free spectral ranges, we can
achieve  single transmission peak in the BPF window, and thus a single
transmission window over the whole optical frequency range. The
transmission profiles shown for the etalons are wider than actual due to the
spectrometer resolution limited at 0.12\,nm. This corresponds to a bandwidth
of about 110\,GHz, so we are unable to resolve the transmission bandwidth of
the etalons. Equally, the
side lobes next to the transmission peak in Fig.~\ref{stack} are an artefact
of the spectrometer we used. 
We estimate the peak transmission of the etalon and filter stack to be around
84\%, improving the transmission of the grating/etalon stack in our earlier
work \citep{pk:14} by about a factor of 4.

The strong emission line at 569.6\,nm in Gamma Velorum ($\gamma$
Vel), which is the brightest Wolf-Rayet star at visual apparent magnitude m$_{v}=1.7$, is a
prime natural laser source candidate \citep{dravins:08}, and would be an 
interesting light source to apply this photon correlation method
to. Therefore, this wavelength was chosen as an instrumental proof-of-concept.
To centre the peak transmission of the etalon stack at 569.6\,nm, we filtered
the blackbody spectrum of an Argon arc lamp with a grating monochromator to a
bandwidth of about  0.12\,nm. With this, we could align the bandpass
interference filters, and tune the etalon temperatures.
The 0.5\,mm etalon had to be maintained at 59.2$^{\circ}$C, and the 0.3\,mm
etalon at 64.0$^{\circ}$C.

\subsection{Intensity correlation scheme}
A Glan-Taylor polariser (GT) selects linearly polarised light to optimise
spatial mode correlations and in conjunction with a polarising beamsplitter
(PBS) and allows to balance the individual detector count rates, thus minimising the measurement duration. This configuration also helps to suppress cross-talk coincidence events arising from the APD breakdown flash \citep{kurtsiefer:01a}.

\begin{figure}
\centering
\includegraphics[width=\hsize]{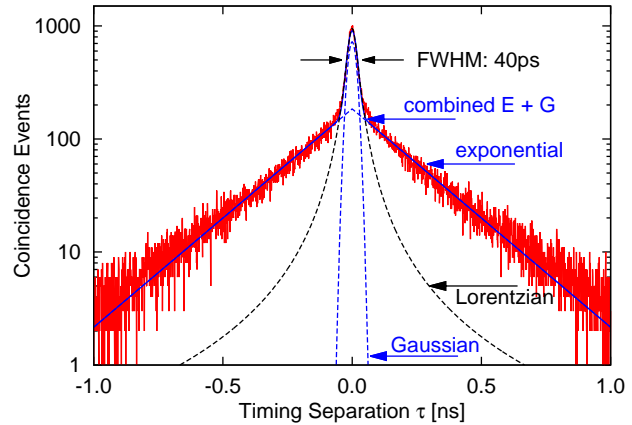}
\caption{Photodetection coincidence histogram from photon pairs at a
  wavelength of 810\,nm generated by parametric down conversion.
  These photon pairs have an intrinsic timing spread of about 0.7\,ps, so this
  coincidence histogram reveals information about the
  timing uncertainty introduced by the detection mechanism only, dominated by
  the APD timing jitter. The dominant central structure can be
  fitted to a Gaussian, with long tails following an exponential decay.}  
\label{mpdtail}
\end{figure}

Finally, the light is detected by two actively quenched Silicon Avalanche
Photon Detectors (APDs). An oscilloscope with a sampling rate of 40\,GS/s was
used to determine the temporal correlation $g^{(2)}(\tau)$. The combined
effective timing jitter of photodetectors and oscilloscope was measured to be
around $\tau_j=40$\,ps (FWHM), but the timing response of the two
photodetectors (taken with a correlated photon pair source based on parametric
down conversion \citep{ling:08}, generating photon pairs around 810\,nm within
a time of about 0.7\,ps, much shorter than the time constants of the
photodetectors) has a peculiar structure (see Fig.~\ref{mpdtail}). The
coincidence distribution of the detectors shows a narrow peak leading to the
small full width at half maximum, but a relatively large base that seems to
reflect an exponential decay. The solid line in the figure represents a
heuristic model 
\begin{equation}
G^{(2)}(\tau)=A\frac{1}{\sqrt{2\pi\sigma^2}}e^{-\tau^2/2\sigma^2}+B\frac{1}{\tau_e}e^{-2|\tau|/\tau_e}\,,
\end{equation}
with a time constant $\sigma=12.0\pm0.1$\,ps for the Gaussian, and
$\tau_e=450\pm2$\,ps for the exponential decay for a fit over the time window $\tau=-1\ldots1$\,ns. The weight $A$ and $B$  of
the two distributions is about the same. We don't have a precise model of the
photodetector, but note that this detector response is most likely what limits
the maximally observed photon bunching signature.

\section{Results}

\begin{figure}
\centering
\includegraphics[width=\hsize]{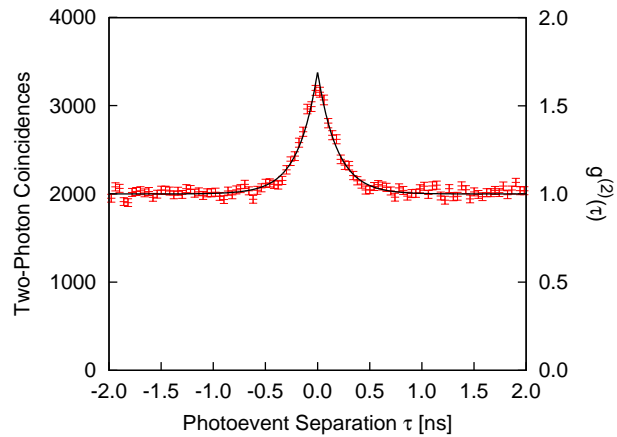}
\caption{Raw coincidence histogram and normalised intensity correlation
  function $g^{(2)}(\tau)$ for the Sun, measured in a 4 minute interval at
  starting 11:36\,am. The characteristic photon
  bunching signature decays exponentially with a time constant of $\tau_c=375\pm35$\,ps
  from the maximum $g^{(2)}(0)=1.69\pm0.05$.}  
\label{g2sun}
\end{figure}

A typical  measurement of the temporal correlation function $g^{(2)}(\tau)$ of
the Sun is shown in  Fig.~\ref{g2sun}. The two-photon coincidence events are
fitted to a distribution
\begin{equation}\label{eq:g2fit}
g^{(2)}(\tau)=1+ae^{-2|\tau|/\tau_c}\,,
\end{equation}
assuming a Lorentzian transmission profile of the etalon stack defining the
frequency distribution, and allowing for a limited interferometric visibility
$V=g^{(2)}(0)/2=(1+a)/2$.  The fitted parameters give a peak value
$g^{(2)}(\tau=0)=1.69\pm0.05$ corresponding to an interferometric visibility
$V\approx$ 85\%, and a coherence time of $\tau_{c}=375\pm35$\,ps.

\begin{figure}
\centering
\includegraphics[width=\hsize]{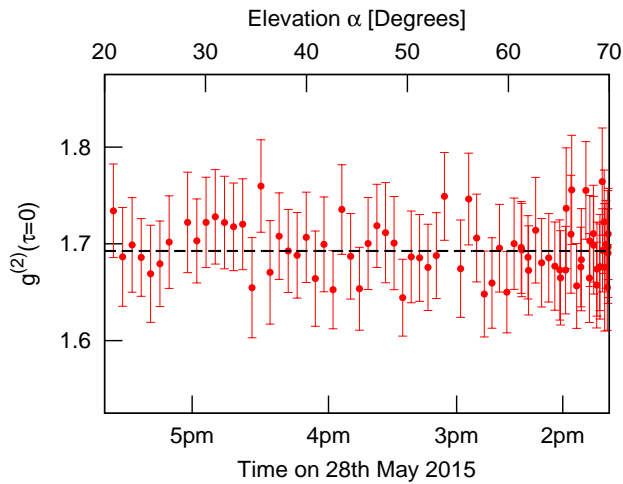}
\caption{Bottom trace: Normalised peak correlation $g^{(2)}(0)$ of the Sun,
  taken in 4  minute intervals over an extended duration from 11:36\,am
  onwards. Top trace: the corresponding elevation $\alpha$ of the Sun.}
\label{trendsun}
\end{figure}

The measurements were conducted in the National University of Singapore observatory site, with geographical coordinates of $1^{\circ}17'49"$N and $103^{\circ}46'44"$E, on the 28th of May 2015, from 11:36\,am to 5:36\,pm (GMT +8:00), through varying cloud cover, weather conditions and elevation angular position of the Sun.
In Fig.~\ref{trendsun}, the Solar $g^{(2)}(\tau=0)$ is shown in 4 minute time
intervals. The corresponding elevation angle
$\alpha$ of the Sun given by
\begin{equation}
  \sin(\alpha-R) = \sin \phi \sin \delta + \cos \phi \cos \delta \cos \omega
\end{equation} 
is shown in the top part of the figure, 
with latitude $\phi$, declination $\delta$ and hour angle $\omega$
\citep{woolf:68} and a small heuristic correction $R$ of at most 34 arc
minutes due to the refraction in the atmosphere  \citep{bennett:82}. 

Over that measurement time where weather and building geometry permitted to
collect data, the elevation angle $\alpha$ covers a range from 
about 70$^\circ$ around noon to 20$^\circ$ in the evening, making the length
of the air
column $s=h_0/\sin\alpha$ the light has to pass through to range from $1.06 h_0$
to $2.92 h_0$, where $h_0$ is the effective height of the atmosphere.
Within the statistical uncertainty, we cannot see any change of the peak
correlation $g^{(2)}(\tau=0)$.

\begin{figure}
\centering
\includegraphics[width=\hsize]{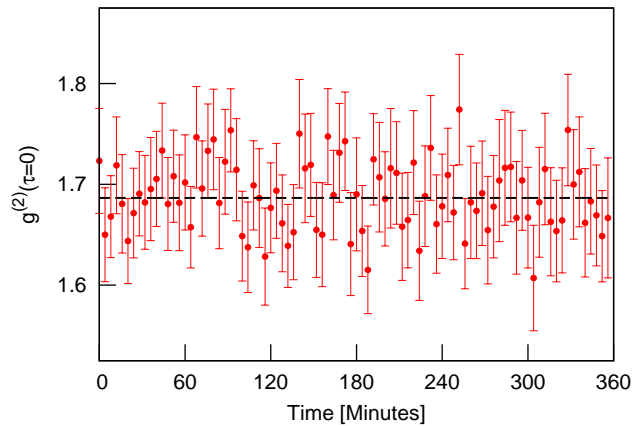}
\caption{Normalised peak correlation $g^{(2)}(0)$ of an Argon arc lamp in
  steps of 4\,minutes for testing the thermal/temporal stability of the
  optical setup.}
\label{trendarc}
\end{figure}

As reference, to investigate for decoherence effects by the atmospheric
turbulence, we compare the observed Solar temporal correlation function  with
one obtained from light of an Argon arc lamp having a blackbody temperature of
6000\,K and thus a suitable analogue to the Sun (see
Figure~\ref{trendarc}), using exactly the same filter configuration as for the
Sun. Both the Sun and the lamp lead to $\approx 10^{5}$
photoevents per second to the APDs after the filtering scheme. For light from
the Arc lamp, we find $g^{(2)}(0) = 1.687 \pm 0.004$. 

The Solar measurement is compatible with a reference measurement performed
with light from an arc lamp.
This agreement suggests that atmospheric turbulence does not degrade the
measurable visibility of temporal intensity interferometry, under our
measurement scheme of 2\,GHz optical bandwidth and 25\,GHz electronic
bandwidth.

The mean Solar $g^{(2)}(0) = 1.693 \pm 0.003$ is significantly higher than our
previous value of $1.37\pm0.03$ \citep{pk:14}, most likely due to the better
off-resonance extinction of the etalon stack, combined with the two
interference filters. The observed distribution
$\tilde{g}^{(2)}(\tau)$ from the filtered Sunlight should be determined by a
convolution of the detector pair response $D(\tau)$ with the idea correlation
function $g^{(2)}$ from (\ref{eq:g2fit}),
\begin{equation}\label{eq:convolution}
\tilde{g}^{(2)}(\tau)=g^{(2)}(\tau)\otimes D(\tau) = \int\!\!g^{(2)}(\tau-\tau') D(\tau')d\tau'\,.
\end{equation}
However, a direct convolution with the normalized measured detector contribution
$G^{(2)}$ shown in Figure~\ref{mpdtail},  
\begin{equation}
D(\tau)=G^{(2)}(\tau)\,/\int G^{(2)}(\tau')\,d\tau'\,,
\end{equation}
significantly underestimates the observed $\tilde{g}^{(2)}(0)$, and leads to a
much larger time constant $\tau_c$ for the exponential decay in
(\ref{eq:g2fit}). We suspect that
the fraction of photodetection events with a long decay time $\tau_e$ seen in
Figure~\ref{mpdtail} depends on the wavelength of the light \citep{ghioni:08},
so a direct application to behaviour at the target
wavelength of 569.6\,nm may not be justified. By using a Lorentzian
distribution with the observed FWHM of $\tau_j=40$\,ps,
\begin{equation}
D(\tau)=\frac{1}{\pi}\frac{\tau_j/2}{\tau_j^2/4+\tau^2}
\end{equation}
as shown in the dashed line in Figure~\ref{mpdtail}, effectively ignoring the
long decay fraction, we obtain a peak value of $g^{(2)}(0)=1.8$, slightly
overestimating the observed distribution. The remaining discrepancy could be
due the deviation of the photodetector response $D(\tau)$ from a Lorentzian,
but also due to minor temperature inhomogeneities in the etalons, or a
difference of the etalon line width due to operating it away from the etalon
design  wavelength of 546.1\,nm.

\section{Discussion and Applications}
There has been a growing interest in recent years to revive the Hanbury-Brown--Twiss method \citep{ofir:06, millour:08,foellmi:09, borra:13, dravins:15}, which has the potential to map the spatial structure of stellar formations \citep{millour:10, dravins:13}, detect exoplanets \citep{hyland:05, strekalov:13} and for kilometric baseline arrays to achieve micro-arcsecond resolution \citep{barbieri:09,borra:13,capraro:09,dravins:07a,lebohec:08,nunez:12}.

Our results suggest no significant decoherence to temporal photon bunching
measurements, thus supporting such proposals. In particular, this measurement
scheme provides for the possibility of temporal coherence studies, which might
be useful in characterising natural laser candidates such as Eta Carinae or
Wolf-Rayet stars \citep{castor:72,hucht:01,johansson:05,varshni:86,roche:12},
or preliminary probes into testing quantum gravity models
\citep{milburn:91,milburn:06,lieu:03,ng:03,ragazzoni:03,maziashvili:09,
  perlman:15}. However, any such decoherence effects would be weak: a
signature of such effects would be a small deviation from the photon bunching
signature with $g^{(2)}(\tau=0)=2$ of a thermal light source. For this, the
current peak value of $g^{(2)}4$ is probably still too low, limited by the
temporal resolution of the avalanche photodiodes.
 
However, with the resolution we have at hand, the Solar $g^{2}(\tau)$ measurements demonstrate its robustness against atmospheric turbulence, affirming intensity interferometry as a serious consideration for both extensive baselines and protracted integration time measurements. 

\section*{Acknowledgements}

We acknowledge the support of this work by the National Research Foundation \&
Ministry of Education in Singapore.




\bibliographystyle{mnras}








\label{lastpage}
\end{document}